\documentclass[aps,pre,twocolumn,showpacs,floatfix,superscriptaddress]{revtex4}

\usepackage{graphicx}
\usepackage{epsfig}

\begin{document}

\title{Activity Dependent Branching Ratios in Stocks, Solar X-ray Flux, and
  the Bak-Tang-Wiesenfeld Sandpile Model} \author{Elliot Martin}\author{Amer
  Shreim}\author{Maya Paczuski} \affiliation{Complexity Science Group,
  Department of Physics and Astronomy, University of Calgary, Calgary,
  Alberta, Canada, T2N 1N4}

\date{\today}

\begin{abstract}
  We define an activity dependent branching ratio that allows comparison of
  different time series $X_{t}$. The branching ratio $b_x$ is defined as $b_x=
  E[\xi_x/x]$.  The random variable $\xi_x$ is the value of the next signal
  given that the previous one is equal to $x$, so $\xi_x=\{X_{t+1}|X_t=x\}$.
  If $b_x>1$, the process is on average supercritical when the signal is equal
  to $x$, while if $b_x<1$, it is subcritical.  For stock prices we find
  $b_x=1$ within statistical uncertainty, for all $x$, consistent with an
  ``efficient market hypothesis''.  For stock volumes, solar X-ray flux
  intensities, and the Bak-Tang-Wiesenfeld (BTW) sandpile model, $b_x$ is
  supercritical for small values of activity and subcritical for the largest
  ones, indicating a tendency to return to a typical value. For stock volumes
  this tendency has an approximate power law behavior.  For solar X-ray flux
  and the BTW model, there is a broad regime of activity where $b_x \simeq 1$,
  which we interpret as an indicator of critical behavior.  This is true
  despite different underlying probability distributions for $X_t$, and for
  $\xi_x$.  For the BTW model the distribution of $\xi_x$ is Gaussian, for $x$
  sufficiently larger than one, and its variance grows linearly with $x$.
  Hence, the activity in the BTW model obeys a central limit theorem when
  sampling over past histories.  The broad region of activity where $b_x$ is
  close to one disappears once bulk dissipation is introduced in the BTW model
  -- supporting our hypothesis that it is an indicator of criticality.

\end{abstract}


\maketitle

\section{Introduction}

Detailed forecasting in complex systems is often difficult if not impossible.
Nonlinear processes as well as long range spatial and/or temporal correlations
can render a direct, reductionist approach futile.  Furthermore, in many cases
of interest, controlled laboratory experiments are unfeasible.  For instance,
stock market data or solar X-ray flux can only be obtained under specific
conditions set by the system itself, and observing the time series under
various controlled conditions is not possible.

Testing the efficient market hypothesis (EMH) presents a clear example of this
difficulty. Roughly speaking, the EMH states that asset prices are inherently
unpredictable~\cite{Samuleson::IndManRev1965, Mandelbrot::JBus1966,
  LeRoy::JEconLit1989}, or that the market is hard to
beat~\cite{McCauley::PhysicaA2008}. There are many flavors of the EMH. The
weak EMH states that the market is efficient if agents only have information
about the time series of market prices. The strong EMH, on the other hand,
states that the market is efficient when agents have access to all relevant
information that could affect prices; this includes e.g. insider trading.

Mathematically the weak EMH can be formulated in terms of a martingale
property for the time series of prices~\cite{Fama::JFin1970}. In its simplest
form, a stochastic variable, $X_t$, is said to be a martingale if the
expectation of its next value given its entire past is equal to its current
value, or $E[X_{t+1}|X_t ...X_0]=X_t$ for all $t$. Empirically, it is not
possible to obtain this expectation value directly from real world time
series.  In fact, the existence of EMH in any of its forms is highly
disputed~\cite{Farmer::PNAS,Lo::JPortMan}.

For a Markov process, the value that $X_{t+1}$ takes only depends on
the previous value $X_t$.  Indeed a necessary but not sufficient
requirement for any stochastic process to be a Martingale is that
$E[X_{t+1}|X_{t}]=X_t$. One example is the critical Galton-Watson (GW)
branching process~\cite{Lyons::2009}. Starting with a single node,
$X_0=1$, each node independently produces a number of offspring that
is  Poisson distributed with mean $b$.  Here $b$ is called the
{\it branching ratio}.  If $b=1$ the process is critical and is also a
Martingale, since $E[X_{t+1}|X_t ...X_0]=E[X_{t+1}|X_t]=bX_t$.

If the underlying probability distribution used to evaluate the
expectation value $E[\cdots]$ is not known, or if the nodes interact
with each other in generating offspring, one can still empirically
measure an {\it activity dependent branching ratio} as $b_x=
E[\xi_x/x]$.  Here the random variable $\xi_x$ is the value of the
next signal given that the previous one is equal to $x$, or
$\xi_x=\{X_{t+1}|X_t=x\}$.  We interpret $\xi_x$ as the set of
outcomes of an interacting branching process with a current population
$x$. Empirically the expectation value $E[\xi_x/x]$ is an average over
all times $t$ when $X_t=x$.  If $b_x=1$ the process is on average
critical when the activity is equal to $x$. If $b_x>1$, it is
supercritical and if $b_x <1$ it is subcritical.  Note that we are not
making any assumptions that the processes we measure are, in fact,
Markovian. The measured branching ratio $b_x$ is an average over all
observed histories leading to a population of size $x$.

We use the activity dependent branching ratio $b_x$ to compare and contrast
time series from stock markets, a physical system (solar X-ray flux) and a
model (the Bak-Tang-Wiesenfeld (BTW) sandpile).  Previously, time series of
activity in the BTW sandpile have been compared in detail with that of solar
flux -- finding a number of similarities~\cite{paczuski::physrevlet2005}. Our
analysis finds similarities as well as significant differences in these two
systems.

We find that $b_x$ is statistically indistinguishable from unity for time
series of stock prices as well as for the Dow Jones industrial average,
consistent with the weak EMH. On the other hand, stock volumes, X-ray flux and
activity in the BTW model all show roughly similar behavior: $b_x$ decreases
from a supercritical value at low levels of activity to a subcritical one at
large values.  This indicates a general tendency for the activity to return to
a characteristic value, which is not present for stock prices.  For stock
volumes, the branching ratio has a relatively strong dependence on activity,
$b_V \sim \left( V/ \langle V \rangle \right)^{-\alpha}$ with $\alpha \simeq
0.69$.  Solar X-ray flux and activity in the BTW model both show a broad range
of activity where the branching ratio is close to one. This broad range
increases with the system size for the BTW model and disappears once bulk
dissipation is introduced, suggesting that it is an indicator of criticality.

We also compare and contrast the probability distributions $P(\xi_x/x)$, at
particular values of activity $x$ in these systems. For the BTW model, this
distribution is well described by a Gaussian for x sufficiently larger than 1.
On the other hand, for both stock volumes and flux intensities, $P(\xi_x/x)$
is broader than Gaussian. The marginal distribution of flux intensities $P(I)$
is well described by a power law, while for the BTW model the distribution of
activity $P(n)$ is approximately exponential with a correlation length that
grows with system size.

In Section II, we present results for the branching ratios determined from
analyses of time series for stock prices and for stock volumes considering
four different stocks as well as the daily Dow Jones average.  Section III
presents results for solar X-ray flux data. Section IV contrasts and compares
results for the canonical BTW model with variants (a) including bulk
dissipation -- making the model subcritical, and (b) having periodic boundary
conditions, which leads to non-ergodic behavior.  We also discuss how our
activity dependent branching ratio differs from the average branching ratio
for avalanches in self-organized critical (SOC) systems previously discussed
in the literature (See
e.g. Refs.~\cite{Carvalho::PhysRevLett,Christensen::PhysRevLett}).  Section V
contains a discussion and summary of the main results.

\section{Stock Market}

Does knowledge of the current price or volume of trade for a particular stock
or market index allow one to make predictions about the next value?  For
prices the answer is no, while for volume of trade the answer is yes, as
discussed next.

We analyze one minute resolution data for four different stocks
from~\cite{HistQuote} for intervals of $28$ days. We also examine one day
resolution data for the Dow Jones over $80$ years from~\cite{YahooF}. Both
price $\xi_P = \{ P(t+1)|P(t)=P \}$, and volume $\xi_V = \{ V(t+1)|V(t)=V \}$
are studied.

\begin{figure}[!h]
  \begin{center}
    \includegraphics*[width=\columnwidth]{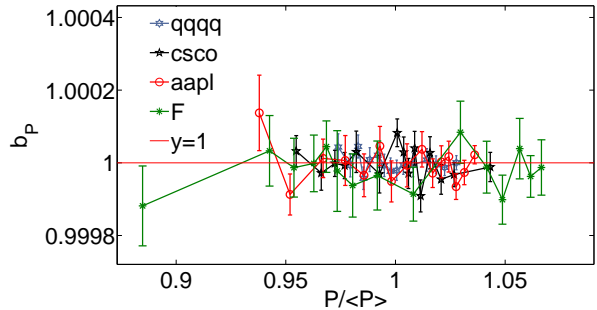}
   \end{center}
   \caption{\label{price-stock} (color on-line) The branching ratio
     $b_p$ vs. price for different stocks.  The x-axis has been scaled
     by the mean price.  All data shown have a resolution of one
     minute, where the price used is that at the start of the
     minute. Error bars in this and subsequent figures indicate one
     standard deviation.  The data for qqqq was taken over the period
     09:30 23/05/08 - 13:37 20/06/08, the data for csco is from 09:30
     23/05/08 - 13:37 20/06/08, the data for aapl is from 9:30
     27/05/08 - 14:08 23/06/08, the data for F is from 09:30 27/05/08
     - 14:03 23/06/08. These results are consistent with the weak
     EMH.}
\end{figure}

\begin{figure}[!h]
  \begin{center}
    \includegraphics*[width=\columnwidth]{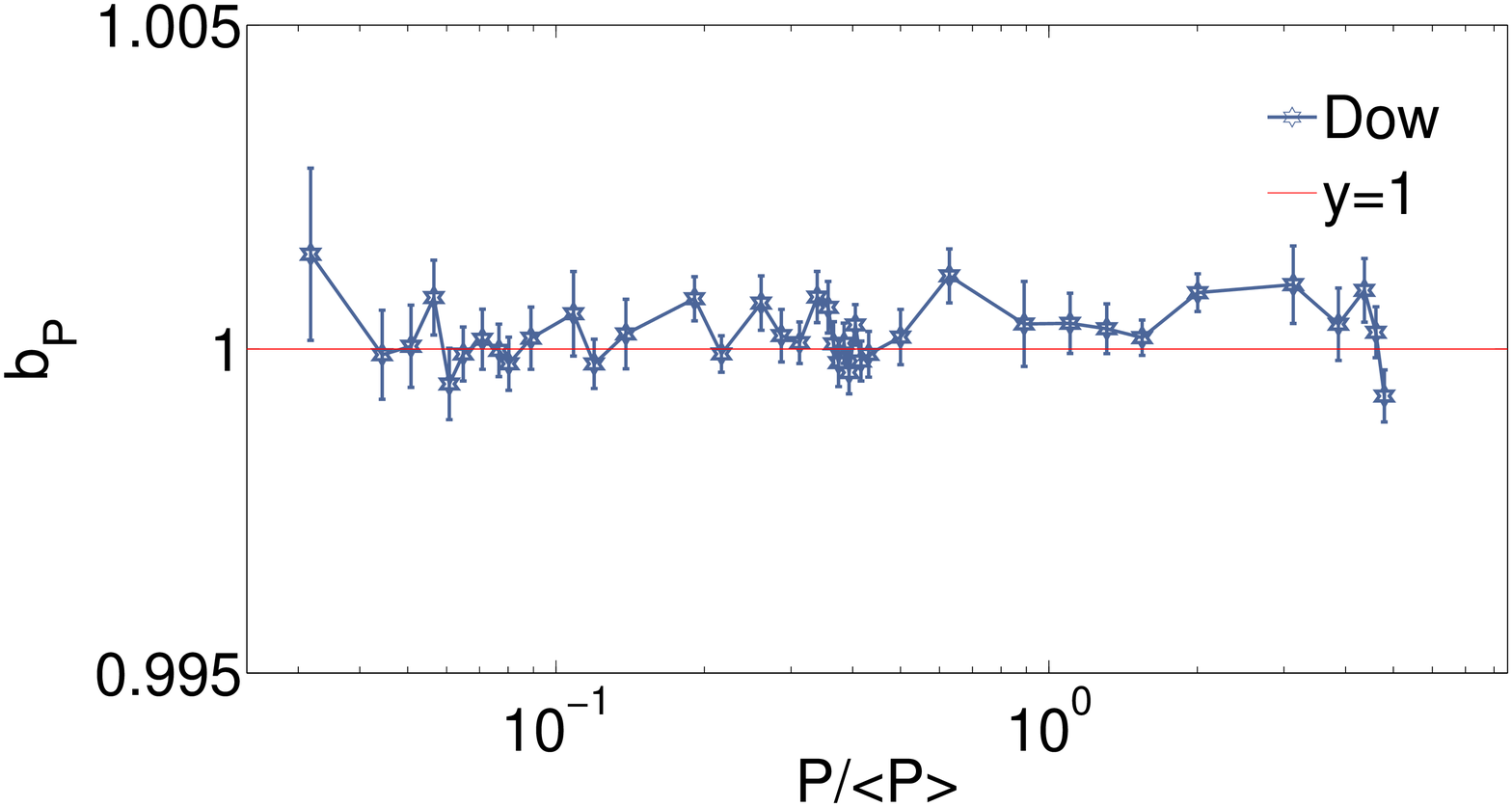}
   \end{center}
   \caption{\label{price-dow} (color on-line) Same as Fig.~1 for the Dow
     Jones industrial average. The data is from
     01/10/1928 - 23/05/2008, and has a resolution of 1 day using the
     opening price. The behavior  is also consistent with the weak EMH.
     }
\end{figure}

Fig.~\ref{price-stock} shows the activity dependent branching ratio
$b_p=E[\xi_p/P]$ vs. $P/\langle P\rangle$ for four different stocks. The same
quantity for the Dow Jones is shown in Fig.~\ref{price-dow}.  The symbol
$\langle\cdots\rangle$ indicates an average over the observation time.  For
all price time series studied $b_p =1$ within statistical uncertainty. These
data are binned such that there are at least 500 points in each bin, and error
bars indicate one standard deviation.

\begin{figure}[!h]
  \begin{center}
    \includegraphics*[width=\columnwidth]{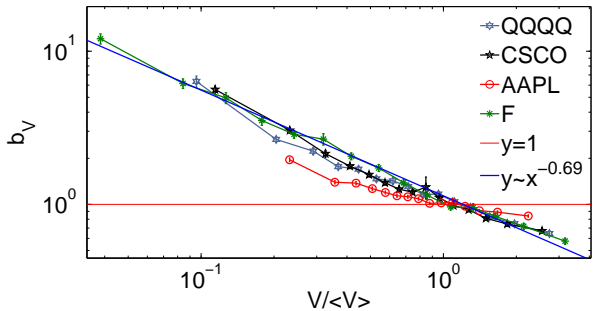}
   \end{center}
   \caption{\label{volume-stock} (color on-line) Activity dependent
     branching ratios for stock volumes during the same time period as
     in Fig.~\ref{price-stock}. A line with slope $m=-0.69$ is
     included as a guide for the eye. The behavior of the Apple stock
     (aapl) appears to deviate from the other three.}
\end{figure}

The activity dependent branching ratio for volume, $b_V$, has a strong
dependence on volume as shown in Fig.~\ref{volume-stock}.  For small values
the stocks behave like a supercritical branching process, while for large
values they are subcritical. Hence the volume has a tendency to return to
roughly its mean value. In fact, $b_V$ has an approximate power law dependence
on $V$, with $b_V \sim \left( V/ \langle V \rangle \right)^{-\alpha}$. The
exponent $\alpha \simeq 0.69$ for three of the stocks shown in
Fig.~\ref{volume-stock} but appears to be smaller (or nonexistent) for the
Apple stock (aapl), which also has a more limited variation in volume,
precluding any firm conclusion about scaling.  We have also analyzed this
quantity for different time windows (data not shown) and the results for the
individual stocks do not vary in any substantial way.  The data for the Dow
Jones volumes (not shown) are too noisy to draw definite conclusions.

\begin{figure}[!h]
  \begin{center}
    \includegraphics*[width=\columnwidth]{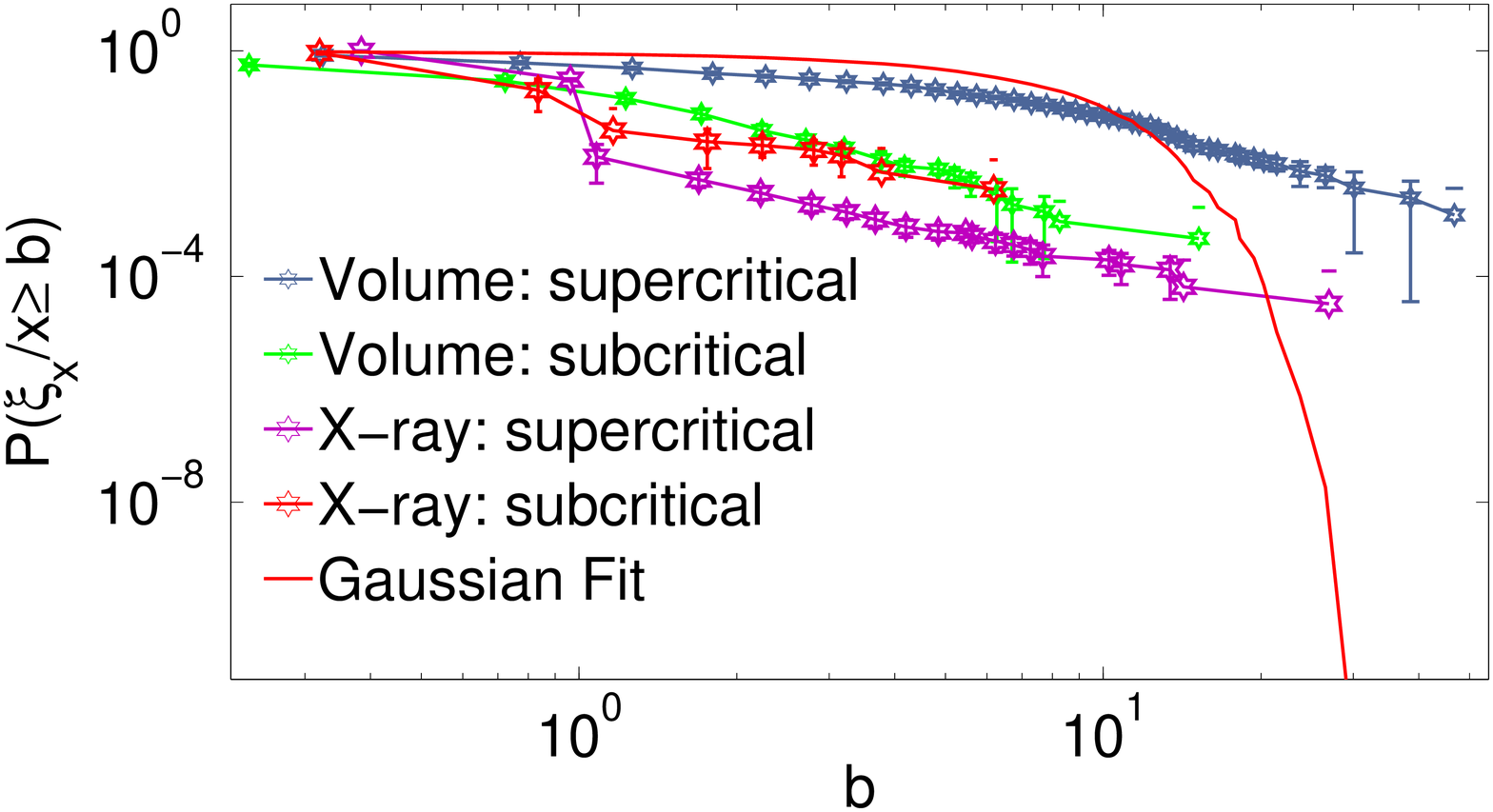}
   \end{center}
   \caption{\label{PDF-compare-vol-sol} (color on-line) Comparison of
     cumulative distribution function (CDF) $P(\xi_x/x \geq b)$ for stock
     volumes and solar intensities at a given value of activity $x$. For stock
     volumes, Ford (F) was used, and the CDFs calculated at $x = V/\langle V
     \rangle = 0.2\pm 0.05$ (supercritical region) and $x = V/\langle V
     \rangle = 2\pm 1$ (subcritical). For solar intensities the entire time
     series, ``All'', was used, and the CDFs calculated at $x = I/\langle I
     \rangle = 0.1\pm 0.005$ (supercritical) and $x = I/\langle I \rangle =
     100 \pm 10$ (subcritical). The Ford CDF at $x=0.2\pm 0.05$ is compared
     with a Gaussian having the same mean and variance. The distributions are
     broader than Gaussian in all cases.}
\end{figure}

For a given level of activity, the probability distribution $P(\xi_V/V)$, also
differs significantly from that for prices $P(\xi_P/P)$.  For prices we find
results similar to that found in Ref.~\cite{Mantegna::Nat}, who analyzed price
changes for different time increments over all prices.  In our case the data
(not shown) are much more noisy since we fix both the initial value of price,
as well as the time interval (one minute).  For stock volumes and solar
intensities, the cumulative distributions $P(\xi_V/V \geq b)$ at a given $V$
and $P(\xi_I/I \geq b)$ at a given $I$, are both broader than Gaussian as shown
in Fig.~\ref{PDF-compare-vol-sol}.

\section{Solar X-ray Flux}

\begin{figure}
\centering
  \includegraphics*[width=\columnwidth]{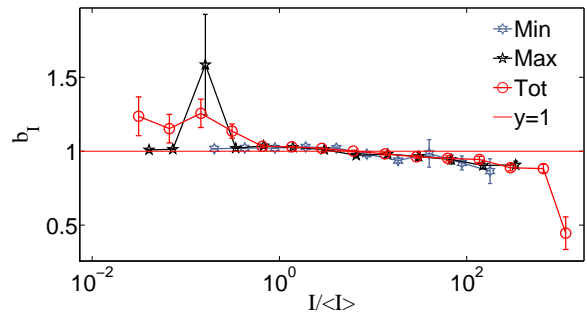}
  \caption{(color on-line) Activity dependent branching ratios for
    solar flux intensity at solar minimum, solar maximum, and for the
    entire data set, ``All'', as defined in the text.  This shows a weak
    tendency to return to a typical value although $b_I\approx 1$ for
    a broad range of intensities, $I$.  This behavior is comparable to that
    shown for the  SOC BTW model in Fig.~\ref{btw-branch}.}
\label{SolExp}
\end{figure}

\begin{figure}[htbp]
  \begin{center}
    \includegraphics*[width=\columnwidth]{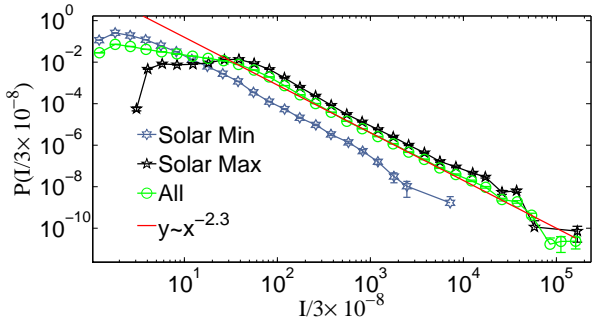}
  \end{center}
  \caption{\label{PDF-sol} (color on-line) The probability
    distribution function of solar x-ray intensities, $P(I)$. The
    straight line indicates a power law fit with exponent $2.3$ for
    the ``All'' time series.  This is different from the approximately
    exponential behavior seen in the SOC BTW model as shown in
    Fig.~\ref{PDF-btw}.}
\end{figure}

Solar flares are bursts of radiation that occur in the solar corona. These
bursts can reach sufficiently high energies to pose a risk to astronauts,
spacecraft, or airplanes following polar routes. In addition they exhibit a
number of empirical features associated with
SOC~\cite{paczuski::physrevlet2005, baiesi::physrevlet2006,
  uritsky::physrevlet2007, Lu::AstrophysJ1991, Paczuski::PhysicaA2004,
  Charbonneau:SolPhys2001}.  For instance, the distribution of event durations
and quiet times is a power-law for both solar flux and the BTW
sandpile~\cite{paczuski::physrevlet2005}, once physically relevant detection
thresholds are taken into account to compare these time series on an equal
basis.  Our analysis shows that the dependence of the branching ratio on
activity are similar in the two cases although the underlying probability
distributions for activity $(X_t)$ and for subsequent conditioned activities
$\xi_x$ are different.

We examine time series in the $1-8$\AA~range obtained from the
Geostationary Operational Environmental Satellites (GOES) Satellites
at the ``Space Physics Interactive Data Resource''~\cite{SPIDR}. A
time series of five minute intervals from 01/01/1986 to 30/04/2008 was
obtained from GOES satellites 5-12. When data from multiple satellites
was available the average of the available data was used. The time
series spans approximately two solar cycles. Data from $10^6-3\times
10^6$ minutes were taken to correspond to a solar maximum, and the
portion from $4\times 10^6 - 6\times 10^6$ minutes to a solar minimum.

Only values of $I>I_0 \geq 3 \times 10^{-8} \mathrm{W/m^2}$ were used when computing
statistics.  This value is close to the detection threshold of the satellites.
In order to make a comparison with the BTW data the solar X-ray intensities
were divided by $I_0$.  This transforms the minimum possible value in both
data sets to one.

The behaviour of $b_I$ is qualitatively similar to $b_V$, as shown in
Fig.~\ref{SolExp}.  The branching ratio decays from supercritical to
subcritical as $I$ increases. However, unlike $b_V$, $b_I$ is close to one
over a broad range of intensities, so the tendency to return to a
characteristic value is much weaker for flux intensities than for stock
volumes. This broad range where $b\approx 1$ is also a property of the SOC BTW
model as shown in the next section. It disappears once bulk dissipation is
introduced into the BTW sandpile; hence we interpret this broad range as a
signal of critical behavior.

Despite this close similarity, the BTW model and solar activity drastically
differ with respect to the distribution of activity, $P(X_t)$, and the
distribution of $(\xi_x/x)$ at a given activity.  As shown in
Fig.~\ref{PDF-sol}, the probability distribution function for flux intensities
$P(I)$ is broad with a tail that is consistent with a power law with exponent
$\simeq 2.3$. On the other hand, the probability distribution of activity
$P(n)$ in the BTW sandpile, shown in Fig.~\ref{PDF-btw}, is close to, but not
exactly, exponential.  In addition the cumulative distribution function
$P(\xi_I/I \geq b)$ at a given level of activity $I$ is broad as indicated in
Fig.~\ref{PDF-compare-vol-sol}.  This contrasts with the BTW model, where the
distribution $P(\xi_n/n)$ at a given level of activity, $n$, is Gaussian (see
Fig.~\ref{PDF-btw-n}).

\section{BTW Sandpile Models}

The BTW sandpile model~\cite{bak::physrevlett1987} is the paradigmatic example
of SOC.  SOC describes slowly driven, dissipative systems that reach a
critical state without fine tuning parameters.  The BTW model depicts a system
that is externally driven to a local dissipative instability whereupon it
``topples''.  This toppling can induce further topplings, which can lead to
cascades of activity propagating through the system. These cascades are called
avalanches. In the steady state, the BTW model reaches a stationary state
where the distribution of avalanche sizes is broad with no natural scale other
than the size of the system~\cite{bak::physrevlett1987,bak::physrevA1988}.

\subsection{Self-organized Critical BTW model}

\begin{figure}[!h]
  \begin{center}
    \includegraphics*[width=\columnwidth]{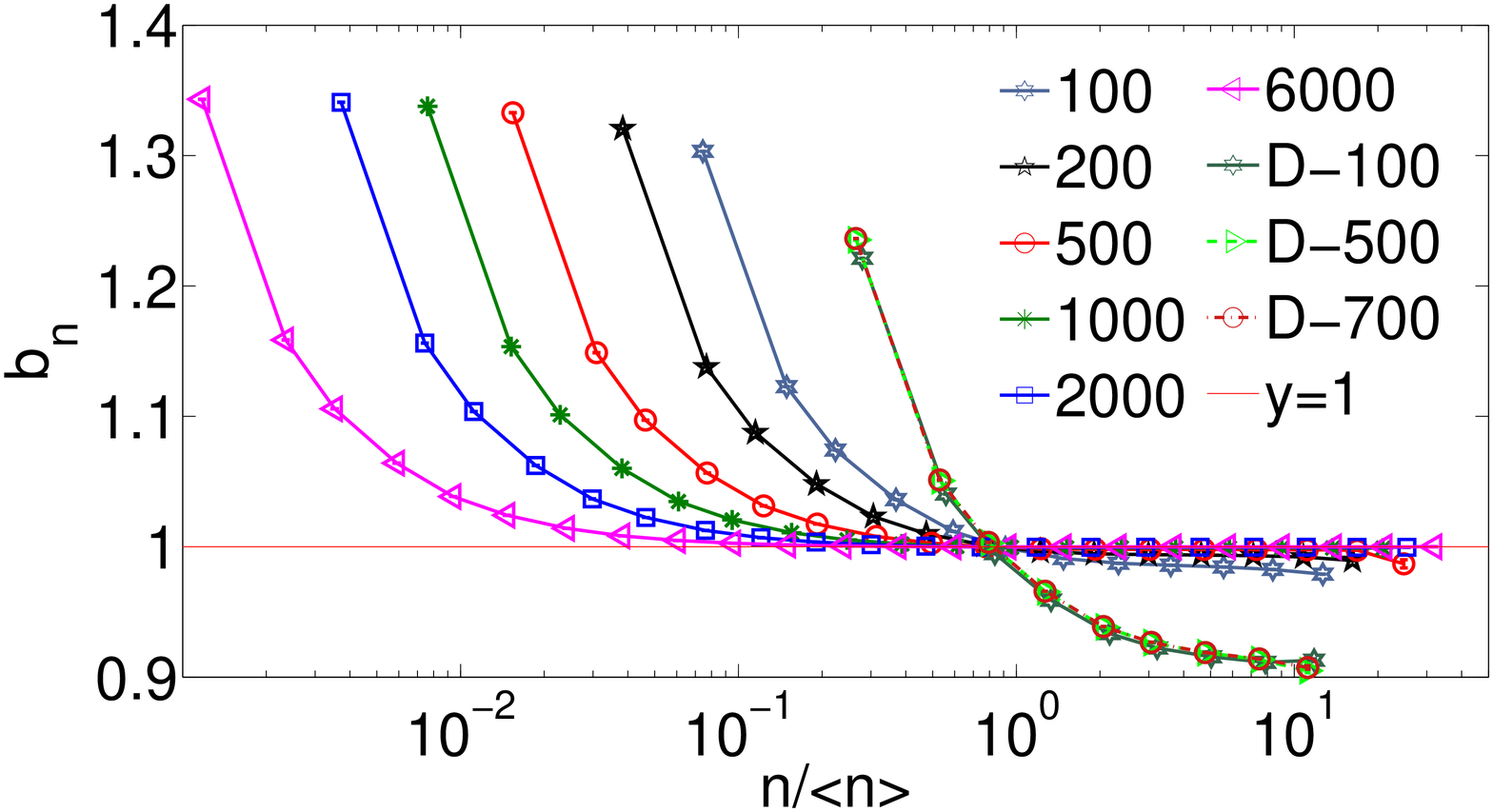}
   \end{center}
   \caption{\label{btw-branch} (color on-line) Activity dependent
     branching ratio for the SOC BTW model for different system sizes,
     $L$. Error bars are smaller than symbol size.  As the system
     size increases, the region where $b_n \approx 1$ broadens.
 The entries beginning with D denote the
     dissipative BTW model.  The dissipative ``D-BTW'' model does not show a
     broad region where $b_n \approx 1$.}
\end{figure}

The SOC BTW sandpile model is composed of an $L\times L$ lattice with open
boundary conditions, where each site is assigned a height $z$. The height of a
stable site is an integer between zero and three. A site with a value $z>3$
becomes unstable and topples by adding a grain to each of its four nearest
neighbors, thus decreasing its height by four. If a boundary site topples, it
throws some grain(s) out of the system.  Initially the sandpile is empty and
$z=0$ for all lattice sites. The system is driven by adding a grain to a
randomly chosen site. Then all unstable sites are updated in parallel, and the
time unit is increased by one. This continues until all sites are stable.
Then another grain is added and the process is repeated {\it ad infinitum}.
We start collecting statistics after the average number of grains in the pile
becomes stationary. A time step corresponds to one parallel update of all
lattice sites, or to the addition of a single grain, whichever is the case.

At every time step, $t$, we record the number of toppling sites, $n_t$.  We
define an activity dependent branching ratio $b_n=E[n_{t+1}/n|n_t=n]$, as the
fraction of sites that topple in a time step immediately following one where
$n$ sites topple. We have numerically simulated the BTW model on lattice sizes
ranging from $L=100$ to $L= 6000$ to study finite size effects.

Fig.~\ref{btw-branch} shows $b_n$ vs. $n/ \langle n\rangle $ for various
system sizes for the SOC BTW model. The behavior is qualitatively similar to
that for solar flare intensities, and for stock market volumes.  The range of
$n$ where $b_n$ is close to one increases with system size $L$.  The
subcritical region occurs for large $n$ as a result of dissipation at the
boundaries, which limits the maximum size of $n$. This means that after a
large toppling event the system is more likely to undergo a smaller one and
$b_n <1$.  We attempted a finite size scaling analysis, which did not give
compelling results. This is consistent with previous results indicating that
the SOC BTW model does not obey finite size
scaling~\cite{Menech::PhysRevE1998, Tebaldi::PhysRevLett1999}.

\begin{figure}[htbp]
  \begin{center}
    \includegraphics*[width=\columnwidth]{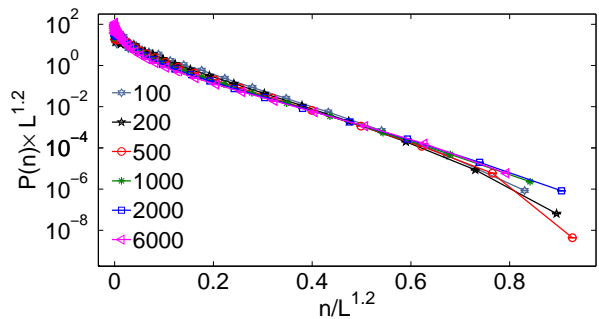}   
  \end{center}
  \caption{\label{PDF-btw} (color on-line) The probability
    distribution for activity in the SOC BTW model, $P(n)$. Error bars
    are smaller than symbol size. The decay is approximately
    exponential.  The best data collapse of the tail of the
    distribution is obtained by rescaling with $L^{1.2}$.  Since the
    SOC BTW model does not exhibit finite size scaling this rescaling
    is only for the purpose of plotting all the data together.}
\end{figure}

Fig.~\ref{PDF-btw} shows the probability distribution $P(n)$, which has an
approximately exponential decay. This differs from the comparable result shown
in Fig.~\ref{PDF-sol} for the solar X-ray intensity. Hence, the behaviour
exhibited by $b_x$ is robust for systems that have markedly different
distributions for activity.  As shown in Fig.~\ref{PDF-btw}, we attempted a
finite size data collapse of the distribution of activity, but this collapse
shows systematic deviations.  However, it is clear that for the distributions
the correlation length increases with system size, leading to a broadening
distribution of activity in the large $L$ limit for the SOC BTW model.

\begin{figure}[!h]
  \begin{center}
    \includegraphics*[width=\columnwidth]
{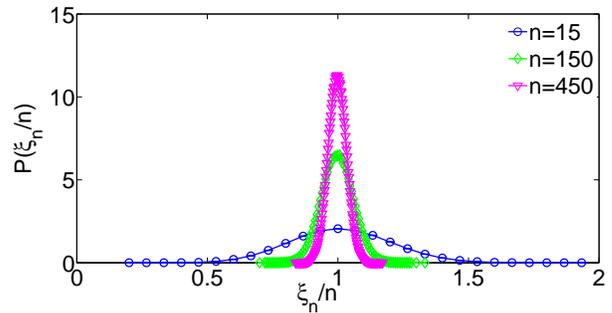}
   \end{center}
   \caption{\label{PDF-btw-n} (color on-line) Probability distribution
     function $P(\xi_n/n)$ for the SOC BTW model. Error bars are
     smaller than symbol size.  The distribution is shown for
     $n=15,150,450$, with $L=500$. The system for $n=15$ lies in the
     supercritical region, while for $n=150$ and $n=450$ it lies in
     the subcritical region.  In all cases the distributions are
     indistinguishable from Gaussian.}
\end{figure}

\begin{figure}[!h]
  \begin{center}
    \includegraphics*[width=\columnwidth]{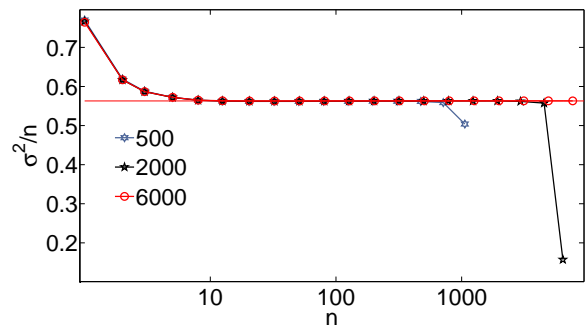}
   \end{center}
   \caption{\label{btw-error-scale} (color on-line) The variance of the random
     variable $\xi_n$, $\sigma^2( \xi_n)/n$ vs.  $n$ for three different
     system sizes. The ratio goes to a constant for large $n$.  Hence the
     variance of $\xi_n$ grows as $n$.}
\end{figure}

We examined the probability distribution, $P(\xi_n/n)$ for various values of
$n$. For $n$ sufficiently larger than one, the distribution is well-described
by a Gaussian.  Fig.~\ref{PDF-btw-n} shows the distribution function for three
values of $n$ in a system of size $L=500$, one in the supercritical regime and
two in the subcritical one.  Fig.~\ref{btw-error-scale} shows that the
variance of the random variable $\xi_n$ increases linearly with $n$.  Gaussian
behavior with a variance that grows linearly with $n$ indicates that the
activity in the BTW model obeys a central limit theorem: when sampling over
prior histories, for each $n$ the activity is the sum of $n$ independent
random processes.

\subsection{Dissipative BTW  Model}

We analyze a BTW model that includes bulk dissipation to test how criticality
affects the activity dependent branching ratio. The model is similar to the
SOC BTW model except it also includes bulk
dissipation~\cite{Vespignani2::PhysRevE}.  When a site topples all its grains
are removed from the system with probability $p_d$, and with probability
$1-p_d$ the normal toppling rule applies.  Fig.~\ref{btw-branch} compares the
branching ratio for the dissipative BTW model with $p_d=10^{-2}$ to the SOC
version. It shows that the broad region where $b_n$ is approximately equal to
one disappears once dissipation is introduced.

\subsection{BTW Model with Periodic Boundary Conditions}
We also studied the BTW model with periodic boundary conditions, so
that no grains are ever thrown out of the system.  As grains are
added, an infinite avalanche eventually occurs. We only examine
statistics of the infinite avalanche. This corresponds to a fixed
energy sandpile, which have been previously studied
in~\cite{Vespignani::PhysRevE,Vespignani::PhysRevLett,Dickman::PhysRevE,Bagnoli::EuroPhysLett}. In
our analysis, an avalanche that lasts more than $9 \times 10^7$
parallel update steps is considered to be infinite, and we only
collect statistics during the infinite avalanche.

\begin{figure}[!h]
  \begin{center}
    \includegraphics*[width=\columnwidth]{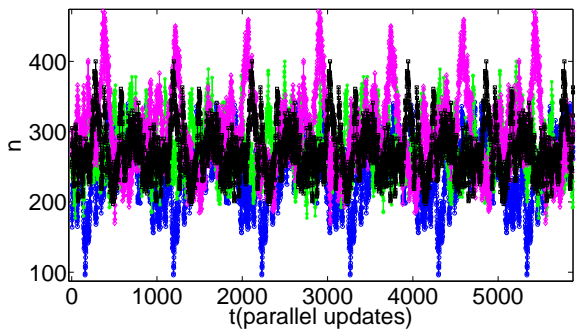}
   \end{center}
   \caption{\label{pb}(color on-line) Time series of activity $n_t$ for
     different initial conditions for the BTW model with periodic boundary
     conditions. This displays different oscillatory behaviour for different
     initial conditions. In each case a system of size $L=500$ was used.}
\end{figure}

Fig.~\ref{pb} shows the time series $n_t$ for  four infinite
avalanches during $6000$ time steps. The figure shows that $n_t$ is
periodic and not ergodic, as was previously noted in
e.g. Ref.~\cite{Bagnoli::EuroPhysLett}. For each realization of the
infinite avalanche the dynamic range of $n$ is small compared to SOC
BTW model.  Moreover, the system is sensitive to
initial conditions. We tested this sensitivity to initial conditions
by starting the lattice empty, or by randomly initializing each site
to a value of $0$ or $1$. The initial conditions affects
 both the period, and the amplitude of oscillations of the
infinite sized avalanche. Similar results were also obtained by simply
keeping the same initial conditions and changing the seed of the
random number generator used. Changing the seed alone was enough to
similarly affect the period and amplitude of oscillations. All these
results imply that the BTW model with periodic boundary conditions is not ergodic and
cannot be compared to the other systems studied here in a meaningful way.

\subsection{Previous definitions of the branching ratio}

The activity dependent branching ratio defined in this paper differs from the
average branching ratio measured
in~\cite{Carvalho::PhysRevLett,Christensen::PhysRevLett}. The previously
defined ratio was not conditioned on activity but rather defined as the
average activity, over all avalanches, resulting from a single
toppling. Indeed it was shown that this average branching ratio ${\bar b}= 1 -
1/\langle s\rangle$, where $\langle s \rangle$ is the average avalanche size.
Hence ${\bar b}$ is not an independent quantity, and is always, by definition,
less than or equal to one, as long as the average avalanche size is finite.
Our activity dependent branching ratio is not restricted to situations where
an avalanche can be well-defined or one can identify individual sites for
activity. In addition it gives an overall picture for how the system behaves
at different levels of activity, unlike the average in
Ref.~\cite{Carvalho::PhysRevLett,Christensen::PhysRevLett}, which sums over
all observed levels of activity.

\section{Discussion and Conclusions}

In this paper we introduced an activity dependent branching ratio, $b_x$, and
use it to analyze different time series, $X_t$, arising in physical, economic,
and model systems.

We found that stock prices have a branching ratio indistinguishable from unity
over for all observed prices.  This observation is consistent with the weak
efficient market hypothesis. Conversely, stock volume, solar X-ray flux, and
the self-organized critical BTW model exhibit supercritical branching ratios
for small levels of activity and subcritical ratios for large ones. This
indicates a tendency for these systems to return to a characteristic
value. This tendency is most pronounced for stock volumes which show a trend
consistent with power law with exponent $\simeq 0.69$, for three out of four
of the stocks examined.  It is not yet clear what separates the Apple stock in
our analysis from the other three, or what the source of the apparent scaling
is.  Solar X-ray flux, and the BTW model both show a broad region where the
activity dependent branching ratio $b_x \approx 1$. When bulk dissipation is
introduced into the BTW model this broad region disappears, supporting our
hypothesis that this is a signature of criticality.

The BTW model and solar X-ray flux show this similarity despite having
different underlying probability distributions for $X_t$.  For solar X-ray
flux the distribution of flux intensities is consistent with a power law with
exponent $\simeq 2.3$, while for the self-organized critical BTW model the
distribution of activity $P(n)$ has an approximately exponential decay, with a
correlation length that grows with system size.

We also found that the variance in activity $\sigma^2(\xi_n)$ scales
linearly with $n$ for the BTW model, and the distribution of
subsequent activity is Gaussian at a fixed $n$.  This indicates that
the BTW model obeys a central limit theorem when sampling over past
histories. It remains to be seen if this last result can be derived
theoretically.

\section{Acknowledgments}
We thank P. Grassberger and V. Sood for useful discussions, and M. Baiesi for
help with the solar flare data.

\bibliography{SOC}

\end{document}